\documentclass{PoS}

\usepackage{amsmath}

\title{Effects of the anomaly on the QCD chiral phase transition}

\ShortTitle{Effects of the anomaly on the QCD chiral phase transition}

\author{\speaker{Shailesh Chandrasekharan} and Abhijit C. Mehta\\
        Department of Physics, Duke University, Durham, NC 27708-0305, USA\\
        E-mail: \email{sch@phy.duke.edu}
}

\abstract{
We study a lattice field theory described by two flavors of massless
staggered fermions interacting with $U(1)$ gauge fields in the strong
coupling limit. We show that the lattice model has a 
$SU(2)\times SU(2)\times U(1)$ chiral symmetry and can be used to
model the two-flavor QCD chiral phase transition in the absence
of the anomaly. It is also possible to add a coupling to this
model which breaks the chiral symmetry to $SU(2)\times SU(2)$
and thus mimics the effects of the anomaly in two-flavor QCD.
We construct an efficient directed loop algorithm to study such
a model. We show that the chiral phase transition in our model
is first order in the absence of the anomaly, while it becomes
second order with $O(4)$ exponents when the anomaly is turned on.
}

\FullConference{XXIVth International Symposium on Lattice Field Theory\\
                July 23-28, 2006\\
                Tucson, Arizona, USA}

\begin{document}

\section{Introduction}

Chiral symmetry plays an important role in determining the low energy physics of QCD. It is also expected to play a vital role in determining the behavior of the chiral phase transition that occurs at finite temperatures. Renormalization group arguments based on $\epsilon$-expansion techniques show that the number of light quark flavors and the strength of the anomaly in QCD can play a significant role in determining the order of the transition \cite{Pisarski:1983ms,Wilczek:1992sf}. While a second order phase transition is possible for two massless quarks, the presence of more light quarks can introduce fluctuations that will force the transition to become first order. If the strength of the anomaly is sufficiently weak at the transition, the two-flavor transition could turn into a first order one. Although $\epsilon$-expansion techniques are not completely justifiable for the problem at hand, there are many examples in which they are known to work sufficiently well. On the other hand there are also examples where these techniques fail.

The effect of the number of light fermion flavors on the chiral phase transition has been a subject of extensive research in the past two decades \cite{Kogut:1982rt,Brown:1990ev,Aoki:1998wg,Karsch:2000kv,Bernard:2004je}. For recent reviews see \cite{Kanaya:1998qw,Laermann:2003cv,Brown:2003wa}. Unfortunately, the effect of the anomaly on the chiral phase transition has not been studied to a similar extent in QCD. Of course it is not possible to tune the anomaly easily in QCD. The subject has been analyzed in the context of mean field theory \cite{Lenaghan:2000kr,Marchi:2003wq} and the results are in agreement with the $\epsilon$-expansion as expected. Thus, if the strength of anomaly at the phase transition was small enough it could be another reason to expect a first order transition in QCD. Lattice simulations have shown that the strength of the anomaly may indeed be small at the chiral phase transition \cite{Chandrasekharan:1998yx,Vranas:1999dg}.

Recently, a new analysis based on a re-summation technique has emerged, which seems to show that the results of the $\epsilon$-expansion in two-flavor QCD in the absence of the anomaly may not be correct \cite{Basile:2005hw}. While the $\epsilon$-expansion forbids a second order transition, the new analysis allows it. It would be interesting to find the predicted second order phase transition and its critical behavior through Monte Carlo methods. Since two-flavor QCD in the absence of the anomaly has an $SU(2)\times SU(2)\times U(1)$ symmetry, the new analysis implies that there is a second order finite temperature critical behavior in the appropriate $O(4)\times O(2)$ sigma model. As far as we know, there are no Monte-Carlo studies in the sigma model context which try to address this question from first principles.

In this work we use strongly coupled two flavor lattice QED with massless staggered fermions to model the physics of the two-flavor QCD chiral phase transition. Although our model is not QCD it has the right chiral symmetries of two-flavor massless QCD and can be studied efficiently with cluster algorithms. Our model may be viewed as an alternative to the usual sigma model approach where symmetries play an important role. Just like in the sigma model we can add a four fermion term to our action which mimics the physics of the anomaly. This allows us to study the phase diagram as a function of the temperature and the strength of the anomaly. The similarity with the sigma model is only true in an indirect sense. More directly, our model is much closer to QCD in spirit since the model is constructed with quark degrees of freedom. Of course the quarks are confined into pions and we can rewrite our model as a statistical mechanics of pion world lines. Thus, our model offers a new method to explain the qualitative features of pion physics of QCD in a simple setting.

\section{Model}

The action of strongly coupled lattice QED that we study is given by
\begin{equation}
S =  - \sum_{x,\alpha} \Bigg\{ \eta_{\alpha,x}
\Big[\mathrm{e}^{i\phi_{x,\alpha}}\overline{\Psi}_{x+\alpha} \Psi_x
- \mathrm{e}^{-i\phi_{x,\alpha}}\overline{\Psi}_{x} \Psi_{x+\alpha}\Big]
\Bigg\} 
- \frac{C}{2} \sum_{x} (\overline{\Psi}_x\Psi_x)^2
\label{lact}
\end{equation}
where $x$ is a lattice site on a four dimensional hyper-cubic lattice with dimensions $L_t\times L^3$. The two component Grassmann fields $\overline{\Psi}_x$ and $\Psi_x$ describe the two flavors of quarks and $\phi_{\alpha,x}$ represents the $U(1)$ gauge field associated to the bond connecting the sites $x$ and $x+\alpha$ ($\alpha=0,1,2,3$ represents the temporal and spatial directions respectively. Further, $\eta_{\alpha,x}$ are phase factors satisfying the relations $(\eta_{t,x})^2 = T$ and $(\eta_{i,x})^2 = 1$ for $i=1,2,3$. We will call the parameter $T$ as the temperature. Thus, we have two tunable parameters $T$ and $C$ in our model.

\begin{figure}
\vskip0.2in
\begin{center}
\includegraphics[width=0.5\textwidth]{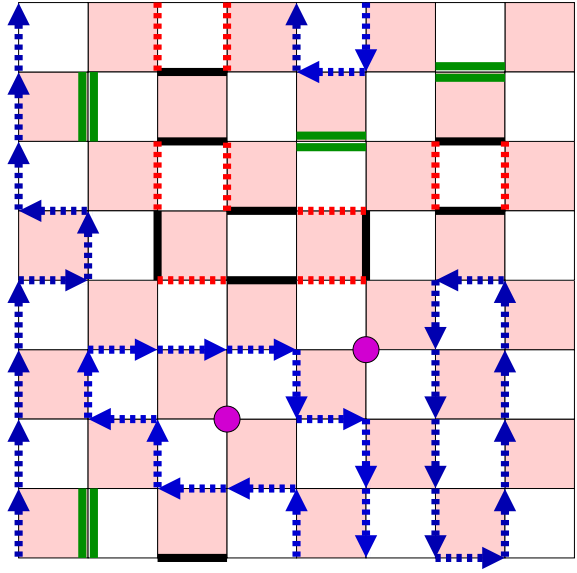}
\end{center}
\caption{\label{fig1} An illustration of a DPI configuration in $1+1$ 
dimensions.}
\end{figure}

It is possible to integrate out the gauge fields completely and rewrite the partition function as a statistical mechanics of Dimer-Pion-loop-Instanton (DPI) configurations. Each configuration is made up of four types of objects: u-dimers $\pi^{u}_{x,\alpha}=0,1$, d-dimers $\pi^{d}_{x,\alpha}=0,1$, oriented pion-dimers $\pi^1_{x,\alpha}=0,1,-1$, and instantons $I_x = 0,2$. A dimer is an object connecting neighboring sites and is associated with the corresponding bond $(x,\alpha)$. There are constraints in the configurations that follow from the fermionic nature of the microscopic action. These can be mathematically represented as follows:
\begin{subequations}
\begin{eqnarray}
I_x + \sum_{\alpha} 
[\pi^u_{x,\alpha} + \pi^d_{x,\alpha} + |\pi^1_{x,\alpha} |
+ \pi^u_{x-\alpha,\alpha} + \pi^d_{x-\alpha,\alpha} 
+ |\pi^1_{x-\alpha,\alpha} |] &=& 0
\\
\sum_{\alpha} [\pi^u_{x,\alpha} - \pi^d_{x,\alpha} 
+ \pi^u_{x-\alpha,\alpha} - \pi^d_{x-\alpha,\alpha}] &=& 0
\\
\sum_{\alpha} [\pi^1_{x,\alpha} + \pi^1_{x-\alpha,\alpha}] &=& 0
\end{eqnarray}
\end{subequations}
The Boltzmann weight of each DPI configuration turns out to be $T^{n_t} I^{n_I}$ where $n_t$ is the total number of temporal dimers (including all the three dimers) and $n_I$ is the total number of instantons. An illustration of a DPI configuration is shown in figure \ref{fig1}

It is easy to verify that when $C=0$ the action is invariant under a $U(2)\times U(2)$ chiral symmetry:
\begin{equation}
\Psi_{x_e} \rightarrow L \Psi_{x_e},\quad \Psi_{x_o} \rightarrow R \Psi_{x_o},
\quad
\overline{\Psi}_{x_e} \rightarrow \overline{\Psi}_{x_e} R^\dagger,\quad
\overline{\Psi}_{x_o} \rightarrow \overline{\Psi}_{x_o} L^\dagger
\end{equation}
where $L,R\in U(2)$ and $x_e$ and $x_o$ refers to even and odd lattice sites. Since the baryon number is gauged, in the strong coupling limit the baryon number is confined. Thus, the relevant symmetry is $SU(2)\times SU(2)\times U_A(1)$ just like the two flavor QCD. When $C \neq 0$, the $U_A(1)$ symmetry is broken, which suggests that it induces the effects of the anomaly. For this reason we call the objects that arise due to $C$ in DPI configurations, as instantons. Our motivation in this work is to study the $T-C$ phase diagram of our model. At low temperature we expect the chiral symmetry to be spontaneously broken, while at high temperatures the symmetry is restored. Our intent is to study the chiral phase transition as a function of $T$. We fix $L_t=4$ in this work.

\section{Observables}

In the low temperature phase we expect $SU(2)\times SU(2)\times U_A(1)$ symmetry to be broken to $SU(2)$ symmetry. In other words $O(4)\times O(2)$ symmetry is broken to $O(3)$. In order to study the nature of the chiral phase transition we focus on the winding number susceptibilities defined using one of the $O(4)$ currents and the $O(2)$ currents.
\begin{subequations}
\begin{eqnarray}
Y_V &=& \frac{1}{L^3} \sum_{\alpha = 1}^3
\Bigg\langle \left( \sum_x  \pi^1_{x,\alpha} \right)^2 \Bigg\rangle
\\
Y_A &=& \frac{1}{L^3} \sum_{\alpha = 1}^3
\Bigg\langle \left( \sum_x  \pi^u_{x,\alpha} + \pi^d_{x,\alpha} + |\pi^1_{x,\alpha}| \right)^2 \Bigg\rangle
\end{eqnarray}
\end{subequations}
We will define $F_\pi^2 = \lim_{L\rightarrow \infty} \chi_V$ and $F^2_\eta = \lim_{L\rightarrow \infty} Y_A$. Both $F^2_\pi$ and $F^2_\eta$ are the squares of the effective three dimensional non-singlet and singlet pion decay constants and should be non-zero when the corresponding symmetries are broken. On the other hand in the high temperature phase when the symmetries are restored we expect the decay constants to vanish. If the chiral phase transition is second order we expect 
\begin{equation}
L Y = f(t L^{1/\nu})
\label{univfn}
\end{equation} 
where $f(x)$ is a universal function analytic at $x=0$ and $\nu$ is a critical exponent, where $t = T-T_c/T_c$ is the reduced temperature.

\section{Results}

\begin{figure}
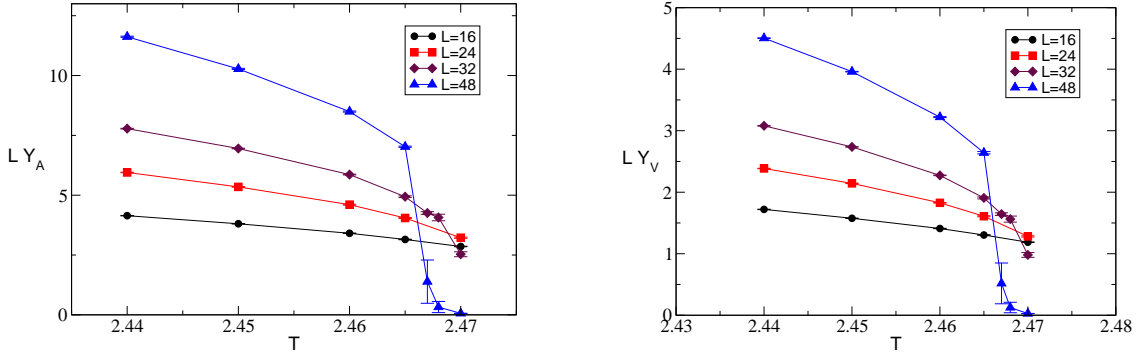

\vskip0.2in
\begin{center}
\hbox{
\includegraphics[width=0.45\textwidth]{fig2a.eps}
\hskip0.5in 
\includegraphics[width=0.45\textwidth]{fig2b.eps}
}
\end{center}
\caption{\label{fig2} Plot of $LY_A$(left) and $LY_V$(right) as a function of $T$ for different lattice sizes at $C=0$. The existence of a second order transition can be confirmed if all the curves intersect at a single value of $T$.}
\end{figure}

When $C=0$ the anomaly is absent. As discussed in the introduction, early studies based on the $\epsilon$-expansion had indicated that the transition must be first order \cite{Pisarski:1983ms}. On the other hand recent studies have suggested the possibility of a second order transition with new critical exponents \cite{Basile:2005hw}. So what does our model show? We have studied this question at $L_t=4$ for lattice sizes $L=24,32,48,64$ and at different temperatures. In figure \ref{fig2} we plot $L Y_A$ (left plot) and $L Y_V$ (right plot) as a function of $T$ for various different values of $L$. The existence of a second order transition can be confirmed if all the curves intersect at a critical temperature $T_c$ as predicted by eq.(\ref{univfn}). Clearly, since the curves do not intersect at a point we conclude that in our model the transition is first order. Our data suggests that $T_c \sim 2.467(2)$.

\begin{figure}
\vskip0.2in
\begin{center}
\includegraphics[width=0.5\textwidth]{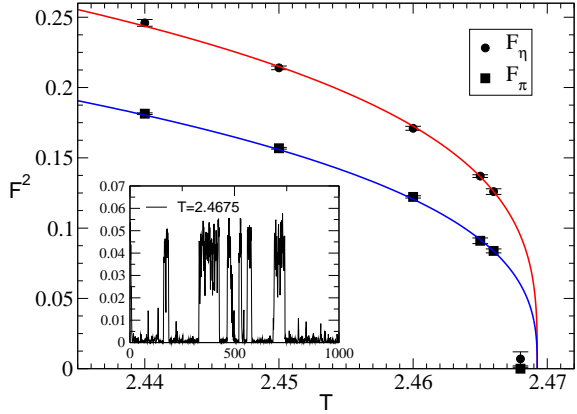}
\end{center}
\caption{\label{fig3} Plot of $F^2_\pi$ and $F^2_\eta$ as a function of $T$. The data fits well to a power law except very close to $T_c$. The inset shows a two state signal in $Y_V$ (as a function of simulation time) at $T=2.4675$ and $L=48$.}
\end{figure}

We have also computed $F^2_\pi$ and $F^2_\eta$ from the large $L$ extrapolations of $Y_V$ and $Y_A$ and these results are plotted in figure \ref{fig3} as a function of $T$. Away from the critical point both $F^2_\pi$ and $F^2_\eta$ fit well to a power-law of the form $A(T_c-T)^\nu$. However, the fit gives $T_c=2.4693(3)$ which is too high, confirming again that the transition is indeed first order. In the inset of figure \ref{fig3} we plot evidence for a two state signal at $T=2.4675$, a value which is close to the critical temperature.

\begin{figure}
\vskip0.2in
\begin{center}
\includegraphics[width=0.5\textwidth]{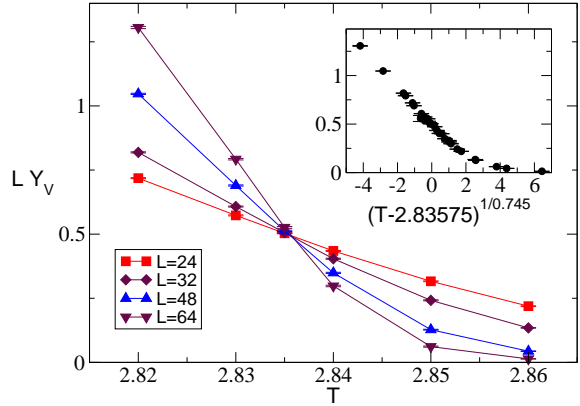}
\end{center}
\caption{\label{fig4} Plot of $LY_V$ as a function of $T$ for different lattice sizes at $C=0.3$. The existence of a second order transition is clearly seen since all the curves intersect at $T_c=2.83575(5)$. The inset shows $L Y_V$ versus $(T-T_c) L^{1/\nu}$ with $\nu=0.745$ expected for a $3d$ $O(4)$ model. }
\end{figure}

Next we switch on the anomaly and study our model at $C=0.3$. In this case the symmetry of our model reduces to $SU(2)\times SU(2)$ which is broken to the diagonal $SU(2)$ at low temperatures. Now the finite temperature phase transition can indeed be second order in the universality class of the three dimensional $O(4)$ spin model. This universality class was studied recently in \cite{ParisenToldin:2003hq} and it was found that $\nu=0.745(2)$. In figure \ref{fig4} we plot $Y_V L$ as a function of $T$ for values of $L$ exactly like figure \ref{fig2}. However, now the curves for different $L$'s do intersect very nicely at a point. In fact we can fit our data, close to the critical point to the form $Y_V L = f_0 + f_1 t L^{1/\nu} + f_2 t L^{2/\nu}$, with $\nu=0.745$. Such a fit reveals that $T_c=2.83575(10)$. In the inset of figure \ref{fig4} we plot $L Y_V$ as a function of $t L^1/\nu$. We see that all of the data collapses nicely to a single function as expected. Thus, we confirm that our model shows $O(4)$ critical behavior at a finite non-zero value of $C$.

In summary, thermodynamics of strongly coupled QED with two flavors of staggered fermions including a four-fermion coupling $C$, which models the chiral phase transition in two flavor QCD with a tunable anomaly, shows a first order transition at $C=0$ and an $O(4)$ critical behavior at $C=0.3$. In particular we do not see the second order transition predicted in \cite{Basile:2005hw}. We are currently trying to locate and study the properties of the tri-critical point present in the $C-T$ plane.

Part of this work was done in collaboration with D.J.Cecile. For a theoretical discussion of the properties of our model and the algorithm used we refer to his contribution to these proceedings. This work was supported in part by a DOE grant DE-FG02-05ER41368.

\end{document}